# Systematic approximations for the period of a finite amplitude pendulum

Ian R. Gatland[a]

*School of Physics, Georgia Institute of Technology, Atlanta, GA 30332-0430*

The standard series expansion for the period of a finite amplitude pendulum as a function of energy (and hence amplitude) provides a lower limit on the period when the series is truncated. An adjustment to the last term in the truncated series to take account of the dropped terms improves the accuracy of the approximation and provides an upper limit on the period. More accurate approximations can then be obtained using intermediate expressions.

## I. INTRODUCTION

Experiments using a pendulum are popular in introductory physics courses as an example of simple harmonic motion but this is only the case when the amplitude is small. Improvements in experimental precision now make small amplitudes impractical. So pendulum experiments require a consideration of finite amplitude periods. Several approximations have been presented for the period of a finite amplitude pendulum with amplitudes up to $90^o$ (a natural limit for a bob on the end of a string).

The most compact approximation is that of Kidd and Fogg[1] who give

$$T = T_0[\cos(\theta_0/2)]^{-1/2}, \qquad (1)$$

where $T_0$ is the period in the small amplitude limit and $\theta_0$ is the amplitude. Equation (1) has a relative error of less than 0.34% for amplitudes up to 75$^\text{o}$ and has a relative error of 0.75% at 90$^\text{o}$. The primary advantage of the Kidd-Fogg formula is that it is very simple but, because of this, it may not be recognized as an approximation.

A somewhat more accurate formula has been proposed, separately, by Molina[2] and Parwani[3] using different derivations. It is

$$T = T_0[\sin(\theta_0)/\theta_0]^{-3/8}. \qquad (2)$$

Equation (2) has a relative error of less than 0.16% for amplitudes up to 75$^\text{o}$ and has a relative error of 0.35% at 90$^\text{o}$.

More recently Lima and Arun[4] have developed the expression

$$T = -T_0 \ln[\cos(\theta_0/2)]/[1 - \cos(\theta_0/2)]. \qquad (3)$$

Equation (3) has a relative error of less than 0.11% for amplitudes up to 75$^\text{o}$ and has a relative error of 0.25% at 90$^\text{o}$.

However, the most accurate of these specialized approximations appears to be the oldest. Over twenty years ago Ganley[5] proposed the formula

$$T = T_0[(\sqrt{3/4}\,\theta_0)/\sin(\sqrt{3/4}\,\theta_0)]^{1/2}. \qquad (4)$$

Equation (4) has a relative error of less than 0.03% for amplitudes up to 75$^\text{o}$ and has a relative error of 0.08% at 90$^\text{o}$. An alternative derivation of Eq. (4) has been given by Parwani[3].

These specialized approximations are convenient but they cannot be directly improved. Also the errors have to be determined by comparison with more accurate calculations or tables.

The exact solution[6] is a complete elliptic integral of the first kind and its standard series expansion converges very slowly. Thus an approximation obtained by truncating the series may require many terms.

However, the approximation based on the truncated series can be improved by an adjustment of the last term. Also, the simple truncated series provides a lower limit on the period and the adjusted version provides an upper limit so the accuracy can be assessed. Further, these two approximations suggest intermediate approximations with significantly better accuracy.

## II. THEORY

Conservation of energy provides an integral expression[7] for the period, $T$, of the finite amplitude pendulum in terms of the amplitude $\theta_0$:

$$T = T_0(2/\pi)\int_0^1[(1-z^2)(1-x^2z^2)]^{-1/2}dz, \tag{5}$$

where $T_0$ is the period in the small amplitude limit and

$$x = \sin(\theta_0/2). \tag{6}$$

The term involving $x$ may be expanded as

$$(1-x^2z^2)^{-1/2} = \sum_{n=0}^{\infty}a_nx^{2n}z^{2n}, \tag{7}$$

where the $a$ coefficients are given by the recurrence relations

$$a_0 = 1 \tag{8}$$

and, for $n > 0$,

$$a_n = [-(1/2-n)/n]a_{n-1} = (1-1/2n)a_{n-1}. \tag{9}$$

Substituting Eq. (7) into Eq. (5) we get

$$T = T_0\sum_{n=0}^{\infty}a_nx^{2n}(2/\pi)\int_0^1z^{2n}(1-z^2)^{-1/2}dz. \tag{10}$$

The integral in Eq. (10) may also be determined by a recurrence relation. Consider the coefficients

$$b_n = (2/\pi)\int_0^1z^{2n}(1-z^2)^{-1/2}dz. \tag{11}$$

For $n = 0$ the integral in Eq. (11) is that for the *arcsin* so we get

$$b_0 = 1. \tag{12}$$

For $n > 0$ we write the integrand in Eq. (11) as $[z^{2n-1}][z(1-z^2)^{-1/2}]$ and integrate by parts to obtain

$$b_n = (2n-1)(b_{n-1} - b_n).$$ (13)

The recurrence relation resulting from Eq. (13) is

$$b_n = (1 - 1/2n)b_{n-1}.$$ (14)

Equations (12) and (14) may also be obtained by noting that the integral in Eq. (11) is a Beta function[6].

By defining the product coefficients

$$c_n = a_n b_n$$ (15)

and combining Eqs. (10), (11), and (15) we have

$$T/T_0 = \sum_{n=0}^{\infty} c_n x^{2n}.$$ (16)

The recurrence relations for the $c$ coefficients are, from Eqs. (8) and (12),

$$c_0 = 1$$ (17)

and, from Eqs. (9) and (14),

$$c_n = (1 - 1/2n)^2 c_{n-1}.$$ (18)

The fact that the $a$ and $b$ coefficients have the same recurrence relations is interesting but coincidental for this study.

The simplicity of the series expansion given by Eqs. (16), (17), and (18) makes it easy to calculate the period to any desired accuracy using a computer program that sums twenty or more terms (quite adequate for amplitudes up to $90^o$ if a relative error of $10^{-7}$ is acceptable).

## III. LIMITING APPROXIMATIONS

The simplest approximation for the period of the finite amplitude pendulum is obtained by truncating the series in Eq. (16) at some $n = m$ so that

$$T/T_0 \cong L_m \equiv \sum_{n=0}^{m} c_n x^{2n}.$$ (19)

Equation (19) is a lower limit because all the terms in the series of Eq. (16) are positive.

An alternative approach is to assume that all the $c$ coefficients are the same for $n \geq m$. Then we have the approximation

$$T/T_0 \cong U_m \equiv \sum_{n=0}^{m-1} c_n x^{2n} + c_m \sum_{n=m}^{\infty} x^{2n} . \qquad (20)$$

Because the $c$ coefficients decrease monotonically Eq. (20) provides an upper limit on the period of the finite amplitude pendulum. Thus Eqs. (19) and (20) bracket the actual period. We can convert the second sum in Eq. (20) to the form

$$\sum_{n=m}^{\infty} x^{2n} = x^{2m} \sum_{n=0}^{\infty} x^{2n} = x^{2m} /(1-x^2) . \qquad (21)$$

Then combining Eqs. (20) and (21) we have

$$U_m = \sum_{n=0}^{m-1} c_n x^{2n} + c_m x^{2m} /(1-x^2) \qquad (22)$$

with the last term in the sum adjusted relative to $L_m$.

Table I compares exact values with those of $L_m$ and $U_m$ for $m = 2$ and $m = 3$. In particular it appears that the relative error in $U_2$ is about 0.1% at 60°, 0.4% at 75°, and 1.3% at 90°. The relative error in $U_3$ is about 0.01% at 60°, 0.08% at 75°, and 0.36% at 90° (but the formula is more cumbersome). In all cases $U$ is more accurate than $L$, but the advantage is not great.

## IV. INTERMEDIATE APPROXIMATIONS

Because $L_m$ is a lower bound and $U_m$ is an upper bound on the period, there is a true value somewhere in between. But its relation to $L_m$ and $U_m$ depends both on $m$ and the amplitude. However, it is worth developing such intermediate approximations. In particular we will consider the form

$$T/T_0 \cong I_m(\alpha) \equiv \sum_{n=0}^{m-1} c_n x^{2n} + c_m x^{2m}/(1-\alpha x^2) \tag{23}$$

with $0 \leq \alpha \leq 1$: $\alpha = 0$ corresponds to $L_m$ and $\alpha = 1$ corresponds to $U_m$. The natural choice is $\alpha = [1 - 1/2(m+1)]^2$ to take care of the next term. The actual choice for the value of the parameter $\alpha$ is a compromise over the amplitudes of interest and is slightly larger.

For the case $m = 1$, $\alpha = 3/5$ gives reasonable results. The explicit formula is

$$T = T_0[1 + x^2/4(1 - 3x^2/5)]. \tag{24}$$

Equation (24) has a relative error of less than 0.035% for amplitudes of less than $75^o$ and a relative error of 0.15% at $90^o$. Thus Eq. (24) is comparable with Eqs. (3) and (4). For the case $m = 2$, $\alpha = 8/11$ gives good results. The formula is

$$T = T_0[1 + x^2/4 + 9x^4/64(1 - 8x^2/11)] \tag{25}$$

and the relative error is less than $1.2 \times 10^{-4}$ (0.012%) for amplitudes up to $90^o$. The next approximation uses $I_3(71/90)$ and has a relative error of less than $2.4 \times 10^{-5}$ for amplitudes up to $90^o$.

## V. COMMENTARY

The actual method used to calculate the period depends on the application and the accuracy required. Equations (1), (4), (24), and (25), and a computer program based on Eqs. (16), (17), and (18) are all viable candidates in appropriate situations.

In any application $U_m$ may be compared with $L_m$ to determine the accuracy of the approximation when the exact value is not known. Thus the appropriateness of the approximation chosen for any particular application can be tested.

Although Eq. (24) is not as compact as Eqs. (1) and (4) for the period of a finite amplitude pendulum, its form suggests that more accurate similar formulae are available. So, even if Eq. (24) is used without understanding its development, the fact that it is one of several possible equations is apparent. This feature is further emphasized if Eq. (25) is compared with the equation for $L_2$ that often appears in textbooks[7].

The approximations discussed here are not only useful in relation to pendulum experiments. They also provide a check for computer programs studying the chaotic pendulum that are developed in computational physics classes.

Table I. Exact periods and limiting approximate periods for a finite amplitude pendulum (relative to the small angle period).

| $\theta_0$ | $L_2$ | $U_2$ | $L_3$ | $U_3$ | Exact |
|---|---|---|---|---|---|
| $15^{\circ}$ | 1.004300 | 1.004301 | 1.004301 | 1.004301 | 1.004301 |
| $30^{\circ}$ | 1.017378 | 1.017423 | 1.017407 | 1.017409 | 1.017409 |
| $45^{\circ}$ | 1.039628 | 1.040145 | 1.039934 | 1.039987 | 1.039973 |
| $60^{\circ}$ | 1.071289 | 1.074219 | 1.072815 | 1.073324 | 1.073182 |
| $75^{\circ}$ | 1.111961 | 1.123332 | 1.116931 | 1.119857 | 1.118959 |
| $90^{\circ}$ | 1.160156 | 1.195313 | 1.172363 | 1.184570 | 1.180341 |